\documentclass[journal]{IEEEtran}

\ifCLASSOPTIONcompsoc
  \usepackage[caption=false,font=normalsize,labelfont=sf,textfont=sf]{subfig}
\else
  \usepackage[caption=false,font=footnotesize]{subfig}
\fi
\usepackage{graphicx}

\usepackage{amsmath,amssymb,mathtools}
\usepackage{bigints}

\usepackage{booktabs}
\usepackage{threeparttable}
\usepackage{tabularx}
\usepackage{multirow}
\usepackage{longtable}
\usepackage{supertabular}
\usepackage{siunitx}

\usepackage{float}
\usepackage{placeins}
\usepackage{flushend}  

\usepackage{cite}
\usepackage{xurl}
\usepackage{textcomp}
\usepackage{enumitem}
\usepackage{fancybox}
\usepackage{adjustbox}
\usepackage{scalerel}
\usepackage[table]{xcolor}

\usepackage{ragged2e} 

\usepackage{algorithm}
\usepackage{algorithmicx}
\usepackage{algpseudocode}
\usepackage{comment}

\algnewcommand{\Inputs}[1]{%
  \State \textbf{Inputs:}
  \Statex \hspace*{\algorithmicindent}\parbox[t]{.8\linewidth}{\raggedright #1}
}
\algnewcommand{\Initialize}[1]{%
  \State \textbf{Initialize:}
  \Statex \hspace*{\algorithmicindent}\parbox[t]{.8\linewidth}{\raggedright #1}
}
\algnewcommand{\AlgInputs}[1]{%
  \State \textbf{Inputs:}
  \Statex \hspace*{\algorithmicindent}\parbox[t]{.8\linewidth}{\raggedright #1}
}
\algnewcommand{\Startingalgorithm}{\State \textbf{Starting algorithm:}}
\algnewcommand{\Endingalgorithm}{\State \textbf{Ending algorithm:}}



\newcommand{\removeFive}{\vspace{-5pt}}

\newcommand{\StepupOne}{\vspace{-\baselineskip}}

\newcommand\doublerulefill{\leavevmode\leaders\vbox{\hrule width .1pt\kern1pt\hrule}\hfill\kern0pt}

\usepackage[hidelinks]{hyperref}
\usepackage{orcidlink}

\begin{document}
\pagenumbering{gobble}
\hyphenation{op-tical net-works semi-conduc-tor}
%
\title{A Novel Piecewise Atmospheric Attenuation Model for Free Space Optical Links in Vertical Heterogeneous Networks}

\author{Eylem Erdogan\textsuperscript{\orcidlink{0000-0003-3657-1721}},~\IEEEmembership{Senior Member,~IEEE,}, Mohammed Elamassie\textsuperscript{\orcidlink{0000-0001-9416-3860}},~\IEEEmembership{Senior Member,~IEEE,}, Ibrahim~Altunbas\textsuperscript{\orcidlink{0000-0001-9416-3860}},~\IEEEmembership{Senior Member,~IEEE,}, Gunes~Karabulut~Kurt\textsuperscript{\orcidlink{0000-0001-7188-2619}},~\IEEEmembership{Senior Member,~IEEE}, Murat~Uysal\textsuperscript{\orcidlink{0000-0001-5945-0813}},~\IEEEmembership{Fellow,~IEEE,}~and~Halim~Yanikomeroglu\textsuperscript{\orcidlink{0000-0003-4776-9354}},~\IEEEmembership{Fellow,~IEEE,}\StepupOne
\thanks{E. Erdogan is with the Izmir Institute of Technology, Turkiye. M. Elamassie is with the Ozyegin University, Turkiye. I. Altunbas is with the Istanbul Technical University, Turkiye. G. K. Kurt is with the Polytechnique Montréal, Canada. M. Uysal is the with the New York University, UAE. H. Yanikomeroglu is with Carleton University, Canada.}}   


\markboth{IEEE Vehicular Technology Magazine}%
{Shell \MakeLowercase{\textit{et al.}}: Bare Demo of IEEEtran.cls for Journals}
%

\IEEEpubid{\begin{minipage}{\textwidth}\centering
  1556-6072  \copyright\,2026 IEEE. All rights reserved, including rights for text and\\ 
  data mining, and training of artificial intelligence and similar technologies.
\end{minipage}} 
\IEEEpubidadjcol

\maketitle

\begin{abstract}
Free-space optical (FSO) communication is emerging as a key backhaul technology for next-generation vertical heterogeneous networks (VHetNets), whose architecture spans satellites, high-altitude platform stations (HAPS), unmanned aerial vehicles (UAVs), and terrestrial nodes. Along these vertical and slant paths, optical beams traverse successive atmospheric layers that may contain clouds, fog, rain, and aerosols, conditions that conventional single-coefficient Beer-Lambert models typically handle only in isolation. Instead of such simplified formulas, we present a unified attenuation model that incorporates aerosols, fog, rain, cloud layers, and drizzle, accounts for the zenith angle, and provides a holistic estimate of the cumulative power loss across atmospheric layers. Numerical results show several-decibel attenuation variations across representative weather scenarios, while the difference between the proposed model predictions and the layer-resolved MODTRAN simulations remains within 1 dB, thereby validating the accuracy of the proposed model and its practical relevance for VHetNet link-budget studies.
\end{abstract}

\begin{IEEEkeywords}
Vertical heterogeneous networks, atmospheric attenuation, Beer-Lambert attenuation model
\end{IEEEkeywords}

\IEEEpeerreviewmaketitle

\section{Introduction}
Vertical Heterogeneous Networks (VHetNets) integrate terrestrial, airborne, and satellite networks into a unified wireless system and are anticipated to play an important role in the development of next-generation wireless systems. In a VHetNet, the layers consist of a satellite (space) network, an aerial network, and a terrestrial network. Low Earth orbit (LEO) mega-constellations comprise multiple satellite orbits and hundreds (often thousands) of small satellites and are expected to bring about seamless global coverage, high robustness, and low latency \cite{VisionMakalesi}. The aerial layer consists of high-altitude platform stations (HAPSs) {in the stratosphere} and unmanned aerial vehicles (UAVs) {in the troposphere}. It is possible to divide HAPS systems into two categories: those that remain airborne using aerodynamic means, such as fixed-wing aircraft, and those that remain aloft using aerostatic means, such as balloons or airships. While the vehicles in the second category can maintain a quasi-stationary position {in the stratosphere}, the HAPSs in the first category rely on continuous movement to stay aloft \cite{HAPS_vision}. Improvements in solar panel technology and lightweight composite materials have enabled the development of aerodynamic HAPS systems, which are expected to play a key role in aerial networks.


For aerial and space backhaul connectivity in a VHetNet, free-space optical (FSO) communication is a promising method for establishing line-of-sight (LOS) links. FSO systems employ laser transmitters with and can offer improved physical-layer security and increased immunity to electromagnetic interference \cite{6844864}. They typically operate within atmospheric transmission windows in the near-infrared and short-wave infrared bands. {For example, commercial FSO terminals commonly operate at wavelengths such as $850$~nm and $1550$~nm, and may also be available at $780$~nm, $1064$~nm, and $1310$~nm} \cite{10.1117/12.417512}. Each wavelength has advantages and disadvantages based on factors such as atmospheric absorption, scattering, turbulence sensitivity, and the availability of components. For example, shorter wavelengths, such as $780$~nm, may be suitable for short-range communications (e.g., where tighter power limits are acceptable), but are generally more susceptible to Rayleigh scattering and solar background noise. In contrast, longer wavelengths such as $1310$~nm and $1550$~nm are better suited for long-range communications due to generally lower Rayleigh scattering, favorable eye-safety and component constraints, and often reduced turbulence-induced scintillation, depending on the operating conditions; however, in heavy fog/haze where Mie scattering dominates, the wavelength advantage may be modest.
\IEEEpubidadjcol

The performance of an FSO communication system is affected by atmospheric molecules and adverse weather conditions. Specifically, particles related to rain, snow, fog, smoke, dust, and aerosols contribute to reduced visibility and high attenuation in signal power. In addition, volcanic-activity-induced sulfate (sulfuric-acid--water) aerosols can also affect the communication quality in the stratosphere \cite{mccormick1993background}. Following major eruptions, these aerosols can persist for months to a few years at altitudes of about $15$--$30$ km above ground level. Hence, optical signals may be significantly attenuated due to the varying contributions of different atmospheric layers to FSO signal attenuation. Understanding these effects requires developing a cascade channel attenuation function that considers the attenuation of each adverse effect.

The attenuation of an FSO system is typically modeled with the Beer-Lambert (BL) formula, which is sufficient for horizontal terrestrial FSO links under nearly constant weather conditions, i.e., a fixed extinction coefficient can be assumed throughout the link distance \cite{6844864}. Many widely used visibility-based formulations (e.g., the Kruse/Kim-type models \cite{Kruse1962}) apply BL attenuation using a single effective extinction coefficient derived from meteorological visibility, which implicitly treats the propagation path as homogeneous. Recent surveys (e.g., \cite{Phuchortham2025}) still summarize and adopt this single-coefficient treatment in performance analyses; however, it is generally not suitable for vertically heterogeneous (slant/vertical) links and motivates an altitude-aware attenuation formulation. In addition to large-scale attenuation, small-scale irradiance fluctuations (turbulence-induced fading) may further affect FSO links; however, such fast fading effects operate on different spatial and temporal scales and can be modeled independently and multiplicatively combined with the proposed large-scale attenuation formulation. Accordingly, for VHetNet slant/vertical links, the classical BL law can be generalized by adopting an altitude-dependent extinction coefficient and integrating it along the propagation path. As an FSO signal propagates upward, it may traverse fog near the ground, rain or snow in the lower troposphere, clouds at various altitudes, and, in some cases, stratospheric aerosol, forming a cascaded attenuation profile along the path.

These atmospheric elements contribute differently to FSO attenuation. Clouds, with their varying densities and altitudes, scatter and absorb parts of the FSO signal. Rain can introduce additional attenuation through absorption and scattering by raindrops, with a dominant dependence on rainfall rate and droplet-size statistics and a secondary dependence on wavelength. Snow particles, which can be larger and more irregular than raindrops, can also cause strong scattering; the net attenuation depends on particle size, wetness, and snowfall rate and may be comparable to (or, in wet-snow conditions, even exceed) rain attenuation. Fog, which can be modeled as a low-altitude cloud layer, further attenuates the signal and can effectively block the link under extremely low visibility \cite{Awan2009}. Additionally, volcanic eruptions can enhance stratospheric extinction by injecting SO$_2$, which is converted to sulfate aerosols (sulfuric-acid droplets) that may persist for months to years. Stratospheric aerosols are generated by major volcanic eruptions. According to the Smithsonian Institution’s Global Volcanism Program (https://volcano.si.edu/), approximately $40$–$50$ volcanoes exhibit ongoing eruptive activity at any given time, while the total number of reported eruptions including both newly initiated and continuing events can reach on the order of $60$–$80$ per year. In this context, a range of $50$–$70$ eruptive events per year is adopted as a representative global benchmark for assessing cumulative stratospheric aerosol loading.

Motivated by this altitude-varying and multi-phenomena attenuation profile, we develop an end-to-end cascade extension of the BL law that partitions the link into altitude slabs, incorporates zenith-angle, and models probabilistic multi-state layer occurrence. This unified formulation goes beyond common treatments that consider these effects separately or rely on a single path-averaged (or piecewise-constant) extinction coefficient, and it is intended to support more accurate link-budget analysis for the space and airborne segments of VHetNets. Our main contributions can be summarized as follows:

\begin{itemize}
\item The proposed piecewise attenuation model can be adopted for slant (accounting for the zenith angle) and vertical VHetNet optical links that traverse the atmosphere, including ground-UAV, ground-HAPS, and ground-satellite links, with varying layer-specific extinction models and vertical extents.
\item In practical scenarios, we model the atmospheric variability within a slab using a {small set of mutually exclusive states}, with occurrence probabilities. 
\end{itemize}


\begin{figure}[!t]
  \centering
    \includegraphics[width=2.8in]{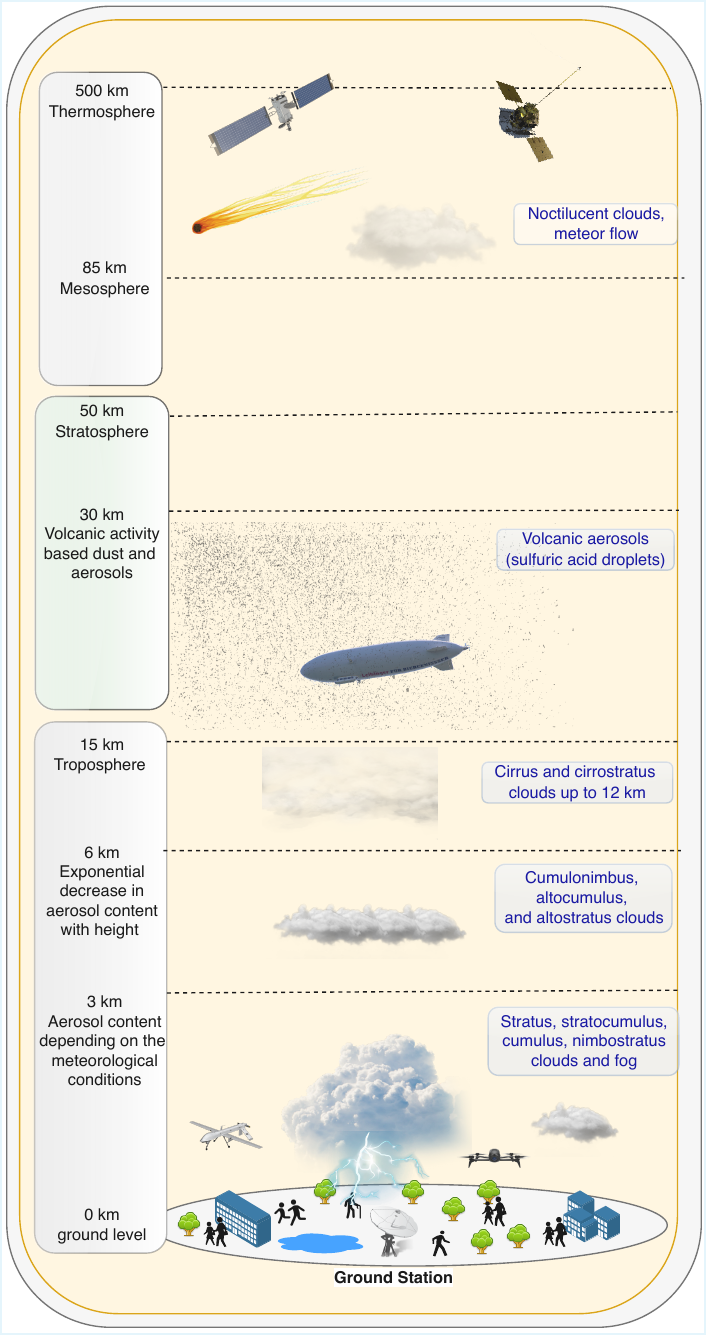}
    \caption{Characteristics of atmospheric aerosols with respect to altitude.}
  \label{fig:model}
\end{figure} 

The remainder of the paper is structured as follows: Section~II describes atmospheric aerosols. In Section~III, we discuss the atmospheric attenuation models. Section~IV considers several scenarios. Within Section~V, we present numerical results, providing our observations. Finally, Section~VI concludes the paper.

\section{Atmospheric Aerosols}
This section summarizes the significant aerosol models with respect to Earth layers as depicted in Fig.~\ref{fig:model}. 

\subsection{Aerosols in Troposphere Layer}

Aerosols refer to solid or liquid particles suspended in air, and they are found throughout the troposphere, from the Earth's surface to the top of the troposphere \cite{sabetghadam2014relationship}. Aerosols within the boundary layer (which is the layer closest to the Earth's surface) have been reported to exhibit the greatest variability \cite{shettle1990models}. {This boundary layer typically extends from the Earth's surface to about $1$ to $2$ km, but it can reach about $3$~km.} However, the composition and properties of aerosols can change rapidly with varying factors such as wind, rain, altitude, and time. This variability makes modeling the first $3$~km of the atmosphere, including the boundary layer, extremely challenging. Additionally, aerosols significantly determine air quality, especially in urban areas, where they contribute to reduced visibility and other environmental concerns \cite{sabetghadam2014relationship}.

\subsection{Aerosols in the Stratosphere Layer}
Aerosols in the stratosphere are not made up of a single, pure chemical compound, nor is their composition stable throughout time or space. After major volcanic eruptions, injected silicate ash can contribute to extinction in the short term, while sulfate aerosols typically dominate the longer-lasting stratospheric attenuation. Furthermore, sulfuric acid solution droplets (often referred to as sulfate aerosols) are widely recognized as the main component of stratospheric aerosols, while ammonium-bearing sulfates may occur under certain conditions. In \cite{elterman1964atmospheric}, it is typically assumed that stratosphere-based aerosols are effective between $15$ to $30$ km height with $15$ km vertical extent as shown in~Fig.~\ref{fig:model}. 


\subsection{Aerosols in Mesosphere and Thermosphere Layers}
The mesosphere ranges from approximately $50$~km to $85$~km in altitude and contains a variety of particulate matter, primarily consisting of meteoric material. These result from meteoroids entering the atmosphere, leading to the formation of particles and debris. In this region and above, tropospheric aerosols such as volcanic ash, pollutants, and dust are generally not sustained, and particulate populations are typically dominated by meteoric smoke particles. In the thermosphere (above about $85$ km), the atmospheric density is extremely low, so particulate aerosols are essentially absent and aerosol-induced optical attenuation is typically negligible. Noctilucent clouds (polar mesospheric clouds) occur near the mesopause (around $82$ to $85$ km). They are composed of ice crystals and are typically observed at high latitudes during the local summer. They can be seen at latitudes between $45^\circ$ N and $80^\circ$ N in the Northern Hemisphere and equivalently in the Southern Hemisphere.

\section{Atmospheric Attenuation Model}
\subsection{Classical Beer-Lambert Attenuation Model}

As light emitted by an optical source traverses the atmosphere, it is progressively weakened by scattering and absorption from gases and aerosols. The classical BL law \cite{6844864} captures this process with an exponential decay, stating that the transmittance is \(\exp(-\omega L)\), where \(\omega\) is the (effective) extinction coefficient and \(L\) is the propagation distance. Practical values of \(\omega\) are typically obtained using radiative-transfer tools such as LOWTRAN or MODTRAN, which can incorporate aerosol and visibility-based extinction models and molecular spectroscopic data (e.g., HITRAN) to compute wavelength-dependent atmospheric attenuation across a wide range of meteorological, latitudinal, and altitudinal conditions; for precipitation effects, especially rain, attenuation is typically parameterized using rain rate rather than visibility.
 For example, the authors in  \cite{9908017}  examined natural aerosol (i.e., not affected by urban or industrial processes) and performed extensive simulations in MODTRAN to determine the extinction coefficients for aerosols within a wavelength range of $350$ nm to $1550$ nm. Accordingly, they developed closed-form expressions applicable to different visibility ranges. In addition to these software programs, Kim's model has been frequently used in BL law to obtain attenuation/extinction coefficient of foggy weathers\cite{10.1117/12.417512}.

In conclusion, the classical BL model is appropriate for horizontal communications where the propagation path can be treated as approximately homogeneous (i.e., dominated by a single attenuation condition with an effective extinction coefficient). For example, consider a foggy scenario where the visibility is very low. In that scenario, the attenuation coefficient (determined by the fog density/visibility) is multiplied by the transmission distance, and the attenuation can be accordingly calculated. Even though the BL model provides accurate calculations in horizontal communications, it cannot be directly used in vertical communications where different atmospheric layers and attenuating conditions exist with different vertical extents.

\subsection{Proposed Piecewise Attenuation Model}
{
Accurate dimensioning of a VHetNet optical link calls for an analytic
attenuation law that tracks the zenith angle \(\zeta\) of a moving
platform and, at the same time, recognizes that different atmospheric
attenuators occupy limited altitude ranges and appear only
intermittently.  The atmosphere is therefore idealized as \(N\)
horizontally uniform slabs; slab \(j\) has vertical extent
\(\Delta L_j\) and, when present, an altitude-averaged extinction
coefficient \(\omega_{j}\,[\mathrm{dB\,km}^{-1}]\).
For the zenith angle \(\zeta\), the slant distance across the
slab equals \(d_j=\sec(\zeta)\,\Delta L_j\), and in this case, the BL law
can be rewritten as 
\begin{equation}
h_j=\exp \left[-\,\sec(\zeta)\,\omega_j\,\Delta L_j\right],
\label{EQN:1}
\end{equation}
where \(h_j\) is defined as the ratio of output to input intensity
after slab \(j\).

Along an entire path, the optical beam encounters specific atmospheric layers probabilistically. This circumstance is captured by an occurrence probability, $\eta_j \in [0,1]$ which quantifies the likelihood that the optical beam traverses the specific atmospheric element (e.g., cloud, fog, or clear air). The overall transmittance then can be obtained as
\begin{equation}\label{EQN:2}
    \begin{split}
        h &= \prod\limits_{j = 1}^N {{h_j}} \\& = \exp \left[ {\sum\limits_{j = 1}^N { - {\kern 1pt} \sec \left( \zeta  \right){\omega _j}} {\kern 1pt} {\eta_j}{\kern 1pt} \Delta {L_j}} \right],
    \end{split}
\end{equation}

\noindent It should be noted that if the extinction coefficient is constant over a given segment, subdividing that segment does not change the predicted attenuation, since $h=\exp\!\left(-\omega(d_{1}+d_{2})\right)=\exp(-\omega d_{1})\,\exp(-\omega d_{2})$. Therefore, sensitivity to the number of layers $N$ arises only when $\omega$ varies with altitude, i.e. $\omega_j\neq\omega_i,~i\neq j$ with $\omega_i$ and $\omega_j$ denoting the extinction coefficients of the $i^{\rm{th}}$ and $j^{\rm{th}}$ layers, respectively, for example, across layer transitions or strong vertical gradients. In practice, the total attenuation is typically dominated by the layer(s) that yield the largest terms $\omega_i \Delta L_i,~i \in \left\{ {1,2, \cdots ,N} \right\}$. Hence, segmentation should focus on capturing these dominant contributors and any sharp vertical gradients, with refinement only where higher-resolution profiles indicate rapid altitude variation.

While (2) provides a compact cascaded expression for the end-to-end transmittance, practical vertical links are affected by atmospheric variability that is inherently stochastic over time and space (e.g., evolving fog intensity, intermittent clouds, or patchy precipitation). In such cases, the same altitude slab can effectively be experienced in different atmospheric states over time (or across nearby realizations). For example, under a semi-clear sky, the link may be subjected to a thick cloud for a certain fraction of time, a thinner cloud for another fraction, and clear air for the remaining fraction. To reflect this graded behavior, we represent slab $j$ by a set of $K_j$ mutually exclusive atmospheric states indexed by $k$, where each state has its own extinction coefficient $\omega_{j,k}$ and occurrence probability $\eta_{j,k}$ that quantifies the long-term fraction of time  for which that state applies. These probabilities satisfy the normalization condition $\sum_{k=1}^{K_j} \eta_{j,k}=1$ for each slab. With this interpretation, the end-to-end transmittance is expressed in the compact probability-weighted form as

\begin{equation}
 h_{\rm{eff}}=\exp \left[-\,\sec(\zeta)
      \sum_{j=1}^{N}\sum_{k=1}^{K_j}
      \eta_{j,k}\,\omega_{j,k}\,\Delta L_j\right],
      \label{EQN:3}
\end{equation}
where $K_j$ is the number of mutually exclusive states considered for slab $j$. For clarity and to keep the model lightweight, the numerical results presented in this work consider up to two representative states per slab, although the formulation itself is not restricted to this choice.

Because each layer is fully specified by the triplet \((\eta,\omega,\Delta L)\), the closed form expression in (3) remains valid for any meteorological regime; by simply substituting the appropriate triplet for volcanic ash, sea salt haze, desert dust, or tropical downpours the same algebra, solution method, and numerical workflow apply without any modification. The vertical limits listed in Table II reproduce the cloud-atlas bands defined in ITU-R P.835-7 \cite{ITUR_P835_7} together with the fog-top statistics used by ICAO, yet the closed form is insensitive to the chosen resolution: shifting any boundary by \(\pm1\) km or splitting a 15 km slab into finer slices changes the 5 km vertical extent loss by less than \(0.3\;\text{dB}\). Users who need higher fidelity can therefore subdivide any layer and append additional \((\eta,\omega,\Delta L)\) rows without altering the mathematics.

}



\begin{table}[!t]
   \renewcommand{\arraystretch}{1.00}
\caption{Atmospheric attenuation and visibility parameters for different atmospheric conditions at $\lambda=1550$ nm.} 
\label{tab1}
\rowcolors{2}{gray!10}{gray!40}
\resizebox{\linewidth}{!}{%
\begin{tabular}{|l|l|l|}
\hline   {Fog} & {Visibility ($V$) km} & { Att. coeff.  (dB/km)  } \\ \hline
\hline  Dense fog  & $0.05$ (low) & $7.0721$ \\
\hline  Thick fog  & $0.20$ (low) & $1.7680$\\
\hline  Moderate fog  & $0.50$ (low) & $0.7072$ \\
\hline  Light fog  & $0.77$ (low) & $0.4592$ \\
\hline \rowcolors{2}{white}{white}  Thin fog & $1.90$ (mod.) & $0.1860$ \\
\hline 
\hline 
{Cloud} & {Visibility ($V$) (km)} & {Att. coeff.  (dB/km)} \\
			\hline \hline
			Cumulus  & $0.0280$ (low) & $12.6287$ \\ 
			Stratus  & $0.0626$ (low) & $5.6486$ \\ 
			Stratocumulus  & $0.0959$ (low) & $3.6872$ \\ 
			Altostratus  & $0.0369$ (low) & $9.5827$ \\ 
			Nimbostratus & $0.0429$ (low) & $8.2425$ \\ 
			Cirrus  & $64.66$ (mod.) & $0.00305$  \\
			\rowcolors{2}{white}{white}
			Thin Cirrus & $290.69$ (high) & $0.00193$  \\
			\hline \hline 
 {Atmospheric poll.} & {Visibility ($V$) (km)} & {Att. coeff.  (dB/km) } \\
			\hline \hline
			Extremely polluted atm. & $1$ (low) & $0.3536$ \\ 
			Normal atm. & $10$ (mod.) & $0.0340$ \\ 
			Non-polluted atm. (clear) & $145$ (high) & $ 0.0025$ \\ 
			\hline \hline

{Snow} & {Visibility ($V$) (km)} & {Att. coeff.  (dB/km) } \\
\hline \hline
{Heavy snow} & {$0.1$ (low)} & {$0.2$} \\ 
			{Moderate snow}  & {$0.5$ (low)} & {$0.08$} \\ 
			{Light snow} & {$1$ (mod.)} & {$0.03$} \\ 
			\hline 
\end{tabular}}
\label{Tab1}
\end{table}
\begin{table}[t]
   \renewcommand{\arraystretch}{1.00}
\caption{Vertical extent, occurrence probability and attenuation coefficient calculations for scenarios under consideration operating at $\lambda = 1550$\,nm.}
\label{notations}
\rowcolors{2}{gray!10}{gray!40}
\resizebox{\linewidth}{!}{%
\begin{tabular}{ |l| l|l| l|  } 
\hline
{{ Scenario} 1 (Rainy weather)} & {Vert. ext.}  & {Occur. prob.} & {Att. Coeff. (dB/km)} \\
\hline
\hline
Nimbostratus & $0.8$ km & $90 \%$  & $8.2425$ \\ 
\hline
Normal atm. & $0.8$ km & $10$\% & $0.034$ \\
\hline
Clear weather & $14.2$ km & $100$\% & $0.0025$ \\
\hline
High volcanic & $15$ km & $50$\% & $0.0104$   \\
\hline
\rowcolors{2}{white}{white}
Background volcanic. & $15$ km & $50$\% & $2.036 \times 10^{-4}$   \\
\hline
\hline
{{ Scenario} 2 (Foggy weather)} & {Vert. ext.}  & {Occur. prob.} & {Att. Coeff. (dB/km)} \\
\hline
\hline
Thick fog & $1$ km & $100 \%$  & $1.7680$ \\ 
\hline
Light fog & $1$ km & $100$\% & $0.4592$ \\
\hline
Clear weather & $13$ km & $100$\% & $0.0025$ \\
\hline
High volcanic act. & $15$ km & $50$\% & $0.0104$   \\
\hline
\rowcolors{2}{white}{white}
Background volcanic & $15$ km & $50$\% & $2.036 \times 10^{-4}$   \\
\hline
\hline
{{ Scenario} 3 (Clear weather)} & {Vert. ext.}  & {Occur. prob. } & {Att. Coeff. (dB/km)} \\
\hline
\hline
Normal atm. & $3$ km & $100 \%$  & $0.034$ \\ 
\hline
Clear weather & $12$ km & $100$\% & $0.0025$ \\
\hline
High volcanic act. & $15$ km & $50$\% & $0.0104$   \\
\hline
\rowcolors{2}{white}{white}
Background volcanic & $15$ km & $50$\% & $2.036 \times 10^{-4}$   \\
\hline
\hline
{{ Scenario} 4 (Extr.\ air poll.)} & {Vert. ext.}  & {Occur. prob.} & {Att. Coeff. (dB/km)} \\
\hline
\hline
Extremely polluted & $3$ km & $70 \%$  & $0.3536$ \\ 
\hline
Normal atm. & $3$ km & $30$\% & $0.034$ \\
\hline
Clear weather & $12$ km & $100$\% & $0.0025$ \\
\hline
High volcanic act. & $15$ km & $50$\% & $0.0104$   \\
\hline
Background volcanic & $15$ km & $50$\% & $2.036 \times 10^{-4}$   \\
\hline
\hline
{{ Scenario} 5 (Snowy weather)} & {Vert. ext.}  & {Occur. prob.} & {Att. Coeff. (dB/km)} \\
\hline
\hline
{Heavy snow} & {$2$ km} & {$100$\%} & {$0.20$} \\
\hline
{Clear weather} & {$13$ km} & {$100$\%} & {$0.0025$} \\
\hline
{High volcanic act.} & {$15$ km} & {$50\%$} & {$0.0104$} \\
\hline
\rowcolors{2}{white}{white}
{Background volcanic} & {$15$ km} & {$50\%$} & {$2.036 \times 10^{-4}$} \\
\hline
\end{tabular}}
\label{Tab2}
\end{table}


\section{Scenarios}
\begin{figure}[!t]
  \centering
    \includegraphics[width=2.8in]{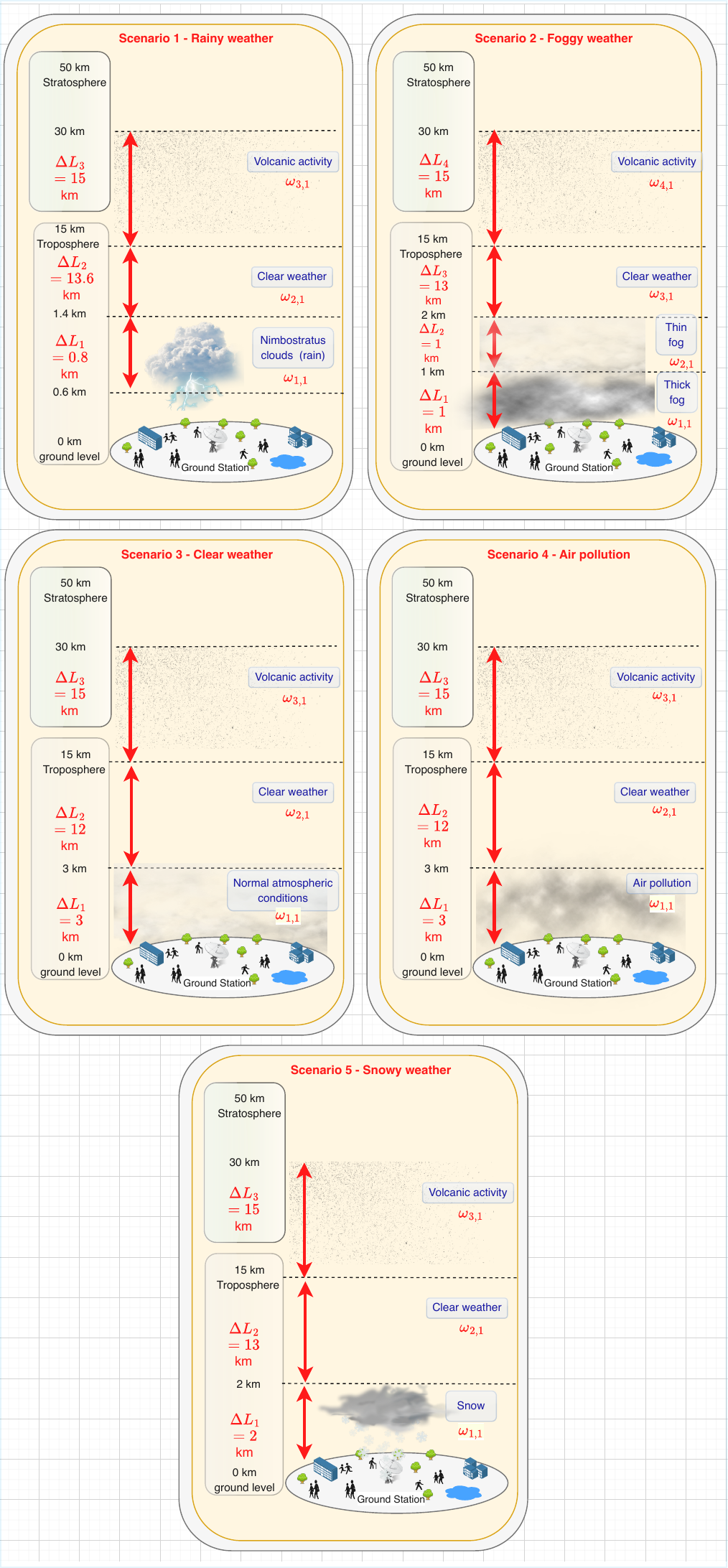}
    \caption{Altitude and vertical extent information about aerosol models in { the considered scenarios}.}
  \label{fig:model2}
\end{figure}

Table \ref{Tab1} shows the visibility parameters and attenuation coefficients for different weather conditions \cite{9908017} where MODTRAN is used to calculate attenuation coefficients based on visibility. In the following, we consider various representative scenarios where we consider the effects of rain, fog, air pollution and compare them with clear weather as shown in Table~\ref{Tab2}. The vertical limits and occurrence probabilities in Table~\ref{Tab2} were data‐driven. Each altitude band follows the standard tropospheric and lower-stratospheric structure prescribed in ITU-R~P.835-7, which sets $0$ – $2$\,km for low cloud, $2$ – $6$\,km for mid-level cloud, $6$ – $12$\,km for high cloud, and $12$ – $25$\,km for the primary stratospheric aerosol reservoir \cite{ITUR_P835_7}, with residual concentrations occasionally extending to $30$\,km following major volcanic eruptions \cite{elterman1964atmospheric}. The $0$ – $0.2$\,km fog cap represents the median fog-top height reported in runway-visual-range statistics, while the $0$ – $10$\,km slab used for background aerosol mirrors the continental-average extinction profile in the same recommendation. Hourly ERA5 re-analysis helped us to calculate the occurrence probabilities: grid points with cloud-fraction~$>0.6$ were counted as rain, liquid-water content~$>0.05$\,g\,m$^{-3}$ as fog, and SO$_2$ column load~$>1$\,DU as volcanic aerosol; normalizing by the total sample count yields the values listed in the Table II \cite{ERA5}. A sensitivity check shows that shifting any layer boundary by~$\pm1$\,km or varying any $\eta_{j,k}$ by~$\pm25\%$ changes the predicted $5$\,km path loss by less than $0.3$\,dB, confirming that the results are robust to reasonable parameter changes. In the calculations, we assume that troposphere has a vertical extent of $15$ km (from $0$ km to $15$ km above the surface level) and the vertical extent of the stratosphere is taken as $35$ km ($15$ km to $50$ km above the troposphere layer). Fig. \ref{fig:model2} shows the scenarios under consideration with altitude, occurrence probability and vertical extent information.

\subsection{{ Scenario} 1: Rainy Weather}
In this first scenario, we consider a rainy day where the weather is associated with nimbostratus clouds which are formed at the lower layer of the troposphere between $0.6$ km to $1.4$ km altitude as shown in Fig. 2  at a vertical extent of $\Delta L_1 = 0.8$ km, bringing rain to the surface. We assume that the optical beam is affected from the frequent nimbostratus cloud formations at an occurrence probability of $\eta_{1,1} = 0.9$, whereas the beam can find a way through normal atmospheric conditions with $\eta_{1,2} = 0.1$ occurrence probability depending on the geographic location of ground stations or transmitter/receiver mobility. Above the cloud formations, the weather is clear up to the edge of the troposphere with a vertical extent of $\Delta L_2 = 14.2$ km and occurrence probability is set to $\eta_{2} = 1$. In the stratosphere, optical beam can be affected either from sulfuric acid ingredients due to high volcanic activity or from standard background stratospheric attenuation \cite{mccormick1993background} at a percentage of $0.5$ occurrence probability with $15$ km ($15$ km to $30$ km height) vertical extent. Based on above values with occurrence probabilities and vertical extent, we obtain the atmospheric attenuation coefficients, as given in Table \ref{Tab2}.


%


\subsection{{ Scenario} 2: Foggy Weather}
In the second scenario, we assume that heavy fog is effective in the troposphere starting from surface level up to $1$ km height (vertical extent $\Delta L_1 = 1$ km). Thereafter, fog level decreases and light fog affects the optical beam for another $1$ km vertical extent up to $2$ km above the surface level as depicted in Fig. 2. After the first $2$ km, clear weather conditions are observed up to the edge of troposphere, and above it, sulfuric acid ingredients causes volcanic activity with same occurrence probabilities as given in Scenario $1$. In this scenario, the occurrence probabilities of heavy fog and light fog are given as $1$. By substituting the attenuation coefficient, vertical extent and occurrence probability into (\ref{EQN:2}), the attenuation caused by foggy weather conditions is obtained.

\subsection{{ Scenario} 3: Clear Weather}
In clear weather conditions, the weather is clear in the first $15$ km starting from sea level up to the edge of the troposphere with an occurrence probability of $1$. In this scenario, we assume non-polluted atmosphere with $V = 145$ km visibility. In the stratosphere, we assume volcanic activity-based sulfuric acid ingredients are effective between $15$ to $30$ km vertical distribution with $0.5$ occurrence probabilities as given in Scenarios $1$ and $2$.

\subsection{{Scenario} 4: Extreme Air Pollution}
In this scenario, we consider an over-populated metropolitan city where the weather is heavily polluted. Thereby, from sea level up to $3$ km altitude, we take air pollution into account. In \cite{sabetghadam2014relationship}, it is observed that $V$ can be $145-225$ km in non-polluted atmosphere, and it can decrease up to $10$ km in normal atmosphere\footnote{In air pollution, all aerosols can contribute to reduced visibility which is a key parameter to precisely calculate attenuation.}. Moreover, in the extremely polluted atmosphere, $V$ can be as low as $1$ km. Therefore, for this scenario, we take $V = 1$ km for a vertical extent of $3$ km with $0.7$ occurrence probability as shown in Table II. The selected occurrence probability is supported by recent large-scale air quality analyses. In particular, the study in \cite{REF:2} reports that more than 70\% of days worldwide exceed the WHO (World Health Organization) daily PM$_{2.5}$ (fine particulate matter) guideline limit, with even higher fractions observed in over-populated regions. This supports the use of a $0.7$ occurrence probability as a representative long-term benchmark for heavily polluted weather conditions. Within the same $0$ - $3$ km altitude range, the beam may alternatively propagate under comparatively less severe atmospheric conditions with an occurrence probability of $0.3$. After the first $3$ km, the weather is clear up to $15$ km (vertical extent is $12$ km) with occurrence probability $1$. In the stratosphere, volcanic activity based sulfuric acid concentration is considered as described in the previous scenarios.

\subsection{ Scenario 5: Snowy Weather}

In this scenario we model a mid-latitude heavy-snow event typical of strong winter cyclones. Radar cross-sections and snow-water-content profiles show that most snow crystals are concentrated in the lowest $2$ km of the troposphere. Visibility can drop from $10$–$15$ km under normal cold-cloud conditions to below $0.5$ km when the snowfall rate exceeds $2$ mm h$^{-1}$. Consistent with this classification, Awan \emph{et al.} \cite{Awan2009} report $100$–$150$ m visibilities for $3$–$5$ mm h\(^{-1}\) events over northern Europe.  Adopting the lower bound \(V = 0.10\;\text{km}\) therefore provides a conservative heavy‐snow benchmark.  Using the visibility–extinction relation derived experimentally at 1550 nm in \cite{9908017}, this corresponds to an attenuation coefficient of \(\gamma_{\text{snow}} = 0.20\;\text{dB\,km}^{-1}\), the value tabulated in Table I. After the snow clouds in $2$ km altitude, the weather is assumed clear up to 15 km (vertical extent 13 km) with occurrence probability 1, identical to Scenarios 1–4.  
In the stratosphere ($15$–$30$ km) we retain the background sulfuric-acid aerosol and episodic high-volcanic layers used earlier, because their optical properties are insensitive to surface snowfall.

\section{Numerical Results}

\begin{figure}[t!]
	\centering
	\includegraphics[width=2.9in]{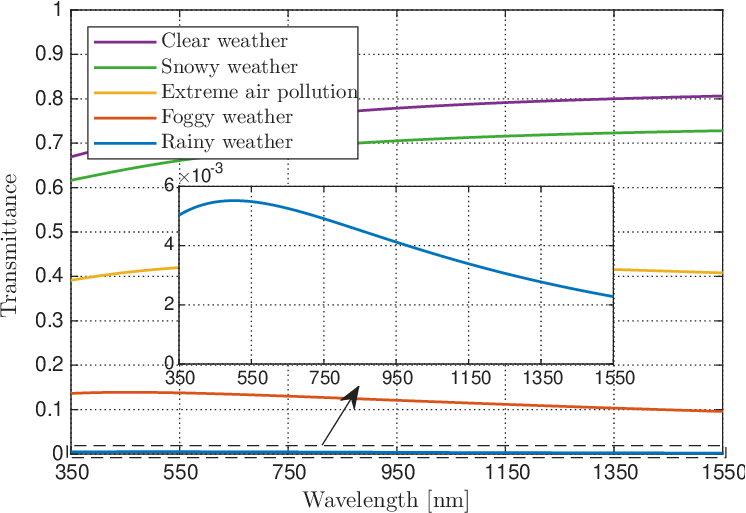}
	\caption{Transmittance vs wavelength for Scenarios $1$, $2$, $3$, $4$ and $5$} when the zenith angle is taken as $\zeta = 10^\circ$.
	\label{fig_3}
\end{figure}
\begin{figure}[t!]
	\centering
	\includegraphics[width=2.9in]{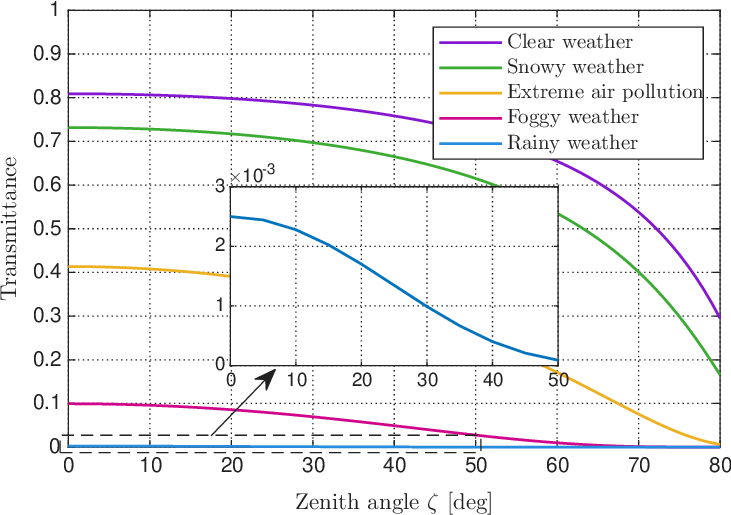}
	\caption{Transmittance vs zenith angle for Scenarios $1$, $2$, $3$, $4$ and~$5$ when the wavelength is set to $\lambda = 1550$ nm.}
	\label{fig_4}
\end{figure}
In this section, we present numerical results for the compact probability-weighted effective transmittance ($h_{\rm{eff}}$) of the scenarios under consideration. Fig. \ref{fig_3} illustrates $h_{\rm{eff}}$ as a function of wavelength, whereas Fig. \ref{fig_4} shows the effect of zenith angle on  $h_{\rm{eff}}$. As can be seen from both figures, the transmittance does not cause a serious problem for optical communication in clear weather, snowy weather as well as in extreme air pollution scenarios as long as $\zeta \leq 40^\circ$. {Both figures further reveal that, in the $1550$ nm window and for zenith angles up to about \(40^{\circ}\), the snow-transmittance curve almost coincides with the clear air baseline; the optical loss remains small because individual ice crystals exhibit negligible absorption at this wavelength.} When the vertical extent of air pollution or occurrence probability increases, extreme air pollution can become a critical problem for vertical communications. Moreover, we also observe that the transmittance of clouds combined with rain affect optical beam much more than foggy weather. This observation can be verified from Table I as cumulus, altostratus, and nimbostratus clouds have higher transmittance coefficients than dense fog. That is why rain clouds (nimbostratus) can increase the optical transmittance up to $2\times 10^{-3}$ at $\zeta = 10^\circ$. Furthermore, we observe from Fig. 3 that atmospheric transmittance does not vary significantly with $\lambda$ for the range of wavelengths under consideration. On the contrary, when zenith angle increases from $\zeta = 40^\circ$ to $\zeta = 80^\circ$, the transmittance rapidly increases as the distance increases with $\zeta$. As a result, communication quality worsens and after $\zeta = 40^\circ$, signal power approximately falls down to one third almost in all scenarios under consideration.


\section{Conclusion}

We have presented a piece-wise BL formulation that merges altitude-segmented extinction, zenith-angle geometry and statistical layer occurrence into a single closed form for vertical and slant FSO links in VHetNets.   By assigning each atmospheric layer its own triplet \((\eta,\omega,\Delta L)\), the model treats clouds, rain, fog, aerosols, and drizzle with their correct depths and probabilities, reproducing layer-resolved MODTRAN results to within about \(0.5\;\mathrm{dB}\).  Attenuation grows almost linearly with \(\sec(\zeta)\), becomes noticeable beyond \(40^{\circ}\), and is dominated by cloud or rain, followed by dense fog, extreme urban haze, and clear air; keeping zenith angles below \(40^{\circ}\) therefore yields a substantial performance benefit.  Because any other weather regime can be represented by substituting its own triplet, rarer phenomena such as dust storms or hurricanes are accommodated by adding a single extra layer, with no change to the algebra or workflow.  Ordinary shifts in visibility, cloud depth, or layer probability move the required fade margin by less than \(1\;\mathrm{dB}\), well inside the \(2\;\mathrm{dB}\) reserve already allotted for pointing error and scintillation, so operators can update margins in real time from a single exponent evaluation, enabling rapid mission planning and on-board power control for UAV and HAPS links.


\section*{Acknowledgments}
	The work of M. Uysal is supported by Tamkeen under the Research Institute NYUAD grant CG017. The work of M.~Elamassie is supported by the Turkish Scientific and Research Council under Grant 125E806.

   $\;$

\bibliographystyle{IEEEtran}

\bibliography{main}

\begin{IEEEbiographynophoto} {Eylem Erdogan} [SM] s a Full Professor at Izmir Institute of Technology and the Director of the Next Generation Sensing and Communication Systems Lab. Previously, he was a Post-Doctoral Fellow at Lakehead University and a Visiting Professor at Carleton University, Canada. His research spans optical wireless and laser satellite communications, non-terrestrial networks, terahertz (THz) communications, physical layer security, and integrated sensing and communications.\removeFive \removeFive\removeFive  
\end{IEEEbiographynophoto}
\begin{IEEEbiographynophoto} {Mohammed Elamassi} [SM]  is an Assistant Professor at Özyeğin University, Istanbul, and Executive Co-Director of OKATEM and the CT\&T Research Group. His research spans optical wireless communications (VLC, FSO, UVLC) and non-terrestrial networks (HAPS, UAV). He has 75+ publications and 1900+ citations (h-index 23). He is an Editor for IEEE Transactions on Communications and an Associate Editor for IEEE Transactions on Vehicular Technology. Honors include the IEEE BlackSeaCom 2019 Best Paper Award and the 2020 IEEE Turkey PhD Thesis Award.\removeFive \removeFive \removeFive\removeFive  
\end{IEEEbiographynophoto}
\begin{IEEEbiographynophoto} {Ibrahim Altunbas} [SM] received the B. Sc., M. Sc., and Ph.D. degrees in electronics and communication engineering from Istanbul Technical University in 1988, 1992, and 1999, respectively. He is currently a Full Professor in the Department of Electronics and Communication, Istanbul Technical University. His current research interests include 6G and beyond wireless communications, satellite/HAPS/UAV communications, index modulation, FSO and THz communications, reconfigurable intelligent surfaces,  ISAC, and data-oriented communications.\removeFive \removeFive \removeFive\removeFive  
\end{IEEEbiographynophoto}
\begin{IEEEbiographynophoto} {Gunes Karabulut Kurt} [SM] is a Canada Research Chairs (Tier 1) in New Frontiers in Space Communications and a Full Professor in Electrical Engineering at Polytechnique Montréal in Montreal. She directs the Poly-Grames Research Centre and is an adjunct research professor at Carleton University. Her research spans 6G and non-terrestrial networks, satellite and aerial communications, and wireless security. She has served as an IEEE Communications Society Distinguished Lecturer and contributes actively to international research and education.\removeFive \removeFive \removeFive\removeFive  
\end{IEEEbiographynophoto}
\begin{IEEEbiographynophoto} {Murat Uysal} [F] is a Professor of Electrical Engineering at New York University Abu Dhabi (NYUAD) and serves as the Founding Director of the NYUAD Wireless Center. Prof. Uysal's research interests are in the broad area of communication theory with a particular emphasis on the physical layer design aspects of wireless communication systems in radio and optical frequency bands.\removeFive \removeFive \removeFive\removeFive  
\end{IEEEbiographynophoto}
\begin{IEEEbiographynophoto} {Halim Yanikomeroglu} [F] is a Chancellor’s Professor in the Department of Systems and Computer Engineering at Carleton University in Ottawa, Canada. He leads the Carleton-NTN Lab and works on 6G and non-terrestrial networks, cellular and radio access architectures, and network-level resource management. An IEEE Fellow (2017), he has served as an IEEE Communications Society Distinguished Lecturer and received the 2024 ComSoc Harold Sobol Award for exemplary service to meetings and conferences in the ComSoc community.\removeFive \removeFive \removeFive\removeFive  
\end{IEEEbiographynophoto}

\end{document}